\documentstyle[12pt]{article}
\def\be{\begin{equation}}
\def\ee{\end{equation}}
\def\bea{\begin{eqnarray}}
\def\eea{\end{eqnarray}}

\author{Hans - J\"urgen Schmidt}

\title{On static spherically 
 symmetric solutions of the Bach-Einstein gravitational field equations}

\date{}
\begin{document}
\maketitle

\centerline{Universit\"at Potsdam, Institut f\"ur Mathematik, Am
Neuen Palais 10} 
 \centerline{D-14469~Potsdam, Germany,  E-mail:
 hjschmi@rz.uni-potsdam.de}

\begin{abstract}
For field equations of  4th order, following from a 
Lagrangian ``Ricci scalar plus Weyl scalar", 
 it is shown (using methods of non-standard analysis) 
that in a neighbourhood of Minkowski space 
there do not exist regular static spherically symmetric solutions. 
With that (besides the known local expansions about $r = 0$ and 
$r = \infty$ resp.) for the first 
time a global statement on the existence of such solutions is given. 
 Finally,  this result will be discussed in connection with Einstein's
particle
 programme. 

\bigskip

F\"ur die Feldgleichungen 4. Ordnung, die aus einem 
Lagrangeausdruck ``Ricci-Skalar plus Weyl-Skalar" folgen, 
wird unter Zuhilfenahme von Methoden der Nicht-Standard-Analysis gezeigt, 
da\ss \   in einer Umgebung
 des Minkowskiraumes keine statischen kugelsymmetrischen 
L\"osun\-gen existieren. Damit wird erstmals neben den bekannten lokalen 
Ent\-wicklungen um $r = 0$ und  $r = \infty$  eine globale Aussage \"uber
 die Existenz solcher L\"osun\-gen getroffen. 
Anschlie\ss{}end wird dies Ergebnis im Zusammenhang mit Einstein's 
Teilchenprogramm diskutiert.
\end{abstract}

\section{Introduction} %1

General Relativity Theory starts from the Einstein-Hilbert Lagrangian 
$L_{\rm EH} =  R/2 \kappa $
 with the Ricci scalar $R$
 which leads to Einstein's vacuum field
 equation $R_i^k= 0$
 being of second order in the metrical tensor $g_{ik}$. Its
 validity is proven with high accuracy
 in space-time regions where the curvature is small only.
 Therefore, the additional presence of a term being 
quadratic  in the curvature is not excluded by the standard weak field
experiments.

Lagrangians with  squared curvature have already been discussed 
by {\sc Weyl}  (1919), {\sc Bach}  (1921), and {\sc Einstein} (1921). 
They were guided by ideas about conformal invariance,
 and  {\sc Einstein}  (1921) proposed to look seriously to such alternatives. 
In {\sc Bach}  (1921)
 $L_{\rm W} = C/2 \kappa $
with the Weyl scalar 
$2C = C_{ijkl}
C^{ijkl}$  was of special interest. 
Then it  became quiet of them for a couple 
of decades because of
 the brilliant results of GRT but not at least because of mathematical
difficulties.

Recently, the interest in such equations has anewed with arguments coming from
 quantum gravity, cf. e.g. {\sc Treder} (1975), 
{\sc Borzeszkowski, Treder, Yourgrau (1978), Stelle (1978), 
Fiedler, Schimming  (1983)}  and ref.
cited there. In the large
 set of possible quadratic modifications a linear combination
$ L = L_{\rm EH} + l^2 L_{\rm W}$  enjoyed a
special interest, cf. e.g. {\sc Treder} (1977). 
The coupling constant $l$  has to be a length for dimensional reasons 
and it has to be a small one to avoid conflicts with 
observations: One often takes Heisenberg length ($ =$  Compton wave
length of the proton) $= 1.3 \cdot 10^{-18}$ cm or Planck length 
$1.6 \cdot 10^{-33}$  cm.

In the present paper we consider this Lagrangian $L$
 in connection with Einstein's particle programme,
 see {\sc Einstein,
 Pauli}  (1943):
 One asks for spherically symmetric singularity-free 
asymptotically flat solutions of  the vacuum field equations 
which shall be interpreted as particles,
  but this cannot be fulfilled within GRT itself; 
  it is still a hope (cf.
 {\sc  Borzeszkowski} (1981))  to realize it in such 4th order field
equations. 
Two partial answers have already been given: 
{\sc Stelle} (1978) showed the gravitationa1 potential of  the linearized 
equations  to be
\be   %1)
\Phi(r) =  -m/r + \exp (-r/l) c/r \quad {\rm for} \quad  r \ll   l \ ;
\ee
he also gave an expansion series about $r = 0$, and
 {\sc Fiedler, Schimming} (1983)
 proved its convergence and smoothness
 in a certain neighbourhood
 of $ r =  0$, but said nothing about the convergence radius of this
expansion.

Now we want to join  these two local expansions. 
To this end we write 
down the field equations in 
different but equivalent versions (sct. 2), calculate some linearizations
(sct. 3) and 
prove the statement of the summary in section 4 
which will be followed by  a discussion of Einstein's particle programme in 
section 5.

\section{Notations and field equations} %2
We start from a space-time metric with signature $(+ - - -)$,
 the Riemann and Ricci 
tensor being 
defined by 
$R^i_{ \ jkl} = \Gamma^i_{ \ jl,k} - \dots$,  
 $ R_{ik} =  R^j_{ \  ijk}$
resp. The Weyl tensor $C^i_{ \ jkl}$
 is the traceless part of the Riemann  tensor and the
Weyl scalar is given by
 $2C = C_{ijkl} C^{ijkl}$. 
 Light velocity is taken to be 1
  and $G = \kappa/8 \pi$
 is Newton's constant; $g = \det g_{ij}$.

 Then we consider the Lagrangian
\be %2
{\cal L}  = \sqrt{-g}  L = \sqrt{-g}  (R + l^2 C)/ 2 \kappa + 
{\cal L}_{\rm mat} \, , 
\ee
where ${\cal L}_{\rm mat}$
 is the matter Lagrangian. 
For writing down the corresponding field equation it is convenient 
to introduce the Bach tensor $B_{ij}$,
  cf. {\sc Bach}  (1921) and {\sc W\"unsch} (1976),
 beforehand.
\be%3
\frac{1}{2} B_{ij} =
 \frac{  \kappa \delta (\sqrt{-g} L_{\rm W})}
{\sqrt{-g} \delta  g^{ij}   }  
  =  C^{a \ \  b}_{\ ij \ ;ba}
 +     \frac{1}{2}     C^{a \ \  b}_{\ ij } R_{ba} \, .
\ee
It holds
$ B^i_i =0$, $B^j_{i;j} =0$, $B_{ij} = B_{ji}$,
  and $B_{ij}$ conformally invariant of weight -1.

Variation 
of  (2)  with respect to $g_{ij}$  leads to the field equation
\be%4
R_{ij} - \frac{1}{2} g_{ij} R + l^2 B_{ij}= \kappa T_{ij} \,  .
\ee
Now let us
 consider the vacuum case $T_{ij}=  0$.
 The trace of eq. (4)  then simply  reads $R=0$, and eq. (4)
 becomes  equivalent to the simpler one
\be%5
R_{ij}  + l^2 B_{ij}= 0 \, .
\ee
Writing $k = l^{-2}$
 and    $\Box \cdot =  g^{ij} \left( \cdot \right)_{;ij} $ 
 one gets also the equivalent system 
\be%6
R = 0 \, ;   \qquad   kR_{ij} + \Box R_{ij} =
 2R_{iabj} R^{ab} + \frac{1}{2}
g_{ij} R_{ab}R^{ab} \, .
\ee
The static spherically symmetric
 line element in Schwarzschild coordinates reads 
\be%7
ds^2 = (1 + \beta) e^{-2 \lambda}
 dt^2 - (1 + \beta)^{-1} dr^2 - r^2d\Omega^2 \, , 
\ee
where $\beta$  and $\lambda$ depend on $r$ only. 
The dot means differentiation with respect to $r$, and defining
$$
\alpha  = 
2 \beta - 2r \dot \beta + 2r \dot \lambda 
(1 + \beta) + 2r^2(\dot \lambda^2 - \ddot \lambda) (1 + \beta) + r^2
(\ddot \beta - 3 \dot \lambda \dot \beta) \, ,  
$$
$$
\zeta = r \dot \alpha \, , \quad \eta_1 = \alpha r^{-3}
 \, , \quad \eta_2 = 3 \beta r^{-1}  \, , \quad \eta_3 = \zeta r^{-3} \, , 
$$
the field equations (6)  are just eqs. (2.15.a-c)
 and (2.16) of the paper of 
 {\sc Fiedler,  Schimming} (1983). To avoid the products
 $ r \dot \eta_i$  we define a new 
independent variable $x$  by $r = le^x$.
 Then  $ r \dot \eta_i  = \eta_i ' =  d\eta_i/dx $ and we obtain
\be%8
0  = \eta_1' + 3\eta_1 - \eta_3 \, , 
\ee
\be%9
0    = (k + r \eta_3 /6) \eta_2 ' + \eta_1 + \eta_3 + r^2k\eta_1
 + r^3 \eta_1\eta_3/6 + r \eta_2\eta_3/2 + r^3\eta_1^2/4 \, , 
\ee
\be%10
0 = (1+r \eta_2/3) \eta_3' - 2 \eta_1 + 2k\eta_2 + 2 \eta_3 - r^2k\eta_1 -
 r^3 \eta_1^2 /2 - r^3 \eta_1\eta_3/6 + 2r \eta_2 \eta_3/3 \, .
\ee
Conversely,  if the system eqs.  (8-10) is solved, the
 metric can be obtained by
\be%11
\beta= r \eta_2/3 \, , \quad \dot \lambda = 
( 2r\dot \eta_2 + 2 \eta_2 + r^2 \eta_1)/(6 + 2r \eta_2  ) \, , 
\quad \lambda(0)=0 \, .
\ee

\section{Linearizations} %3
Now we make some approximations to obtain the rough 
behaviour of the solutions. 
First,  for $r \to \infty$, 
{\sc Stelle}  (1978)
 has shown the following: Linearization about Minkowski 
space\footnote{That means, loosely
 speaking, the right hand side of eq. (6) will be neglected. 
Connected with this one may doubt the relevance of the term 
$cr^{-1}e^{-r/l}$  in eq. (1) for
 $m \ne  0$ and $r \to \infty$,  because it is small compared with the 
neglected terms.}
 leads (using our notations (1) and (7)) to $\lambda 
 = 0$ and $\beta =   2 \Phi$,   where powers of $\Phi$
 are neglected and the term  $\exp (r/l)/r$ in $\Phi$
 must he suppressed because of asymptotical flatness.
 Further, a finite $\Phi(0)$  requires $c = m$  in (1). 
Then $\Phi$  is just the Bopp-Podolsky 
 potential (stemming from 
4th order electrodynamics),
 and it were {\sc Pechlaner, Sexl}  (1966)
 who proposed this form as representing a gravitational potential.

Now we have 
$\Phi(r) = m \left( \exp (-r/l) - 1 \right)/r = -m/l + rm/(2l^2) + \dots $, 
but also this finite potential 
gives rise to a singularity: 
the invariant $R_{ij}R^{ij} \approx m^2/r^4$ as $r \to  0$,
 that means, the linearization makes no sense for this region. 
From this one can already expect that in a neighbourhood 
of Minkowski space no regular solutions exist.

Second, for $r \to  0$,
 {\sc  Fiedler, Schimming} (1983) 
proved that there exists a one-parameter family 
of solutions being singularity-free and analytical in 
a neighbourhood of $r = 0$. 
 The parameter will be called $\epsilon$
 and can be defined as follows: neglecting
 the terms with $r$
 in eqs. (8-10)
 one obtains a linear system with constant coefficients possessing 
 just a one-parameter family of solutions being regular at $r = 0$; it reads
\be%12
 \eta_1 = \epsilon k^2r \, , \quad
\eta_2 = -5\epsilon k r \,  ,   \quad   \eta_3  = 4 \epsilon k^2r \,  .   
\ee
(The factors $k^n$
 are chosen such that $\epsilon$
 becomes dimensionless.) Now one 
can take (12) as the first term of a power series 
 $\eta_i = \Sigma_n \, a_i^{(n)} r^n$ , 
and inserting this 
into eqs. (8-10) one iteratively obtains the coefficients $a_i^{(n)}$.
For $n$ even, $a_i^{(n)}$
 vanishes:
 The 
$r^3$-terms are $k^3(\epsilon/14 + 10\epsilon^2/21)$,
 $ -k^2(\epsilon/2 - 10\epsilon^2/3)$  and 
$k^3(3\epsilon/7 + 20\epsilon^2/7)$ resp. 
Furthermore,
 the $r^{2n-1}$-term is always a suitable power of $k$
 times a polynomial in $\epsilon$ of the 
order $\le n$.

Up to the $r^2$-terms the corresponding metric (7) reads
\be%13
     ds^2 = (1 + 5k\epsilon r^2/3) dt^2
 - (1 + 5k\epsilon r^2/3) dr^2 - r^2d \Omega^2 \, .
\ee

Third, we look for 
 a linearization which holds uniformly for $0 \le  r < \infty$. 
A glance at (11)
 and (7) shows that $\eta_i = 0$
 gives Minkowski space, and therefore we neglect
 terms containing products of  $\eta_i $  in (8-10).
 In other words,
instead of the linearization 
used before we additionally retain the term $r^2 \eta_1$.  
Then we proceed as follows: from eqs. (8)
and (10) we obtain
\be%14
\eta_3 = \eta_1' + 3 \eta_1      \qquad {\rm  and}
\ee
\be%15
\eta_2 =\frac{- \eta_1   ' - 5 \eta_1   + (e^{2x} - 4) \eta_1
+ l^3 e^{3x} \eta_1
(\eta_1 + \eta_1' /6)
}{2k+ l e^x (\eta_1'   + 5 \eta_1   + 6
\eta_1)/3} \, .
\ee
Inserting eqs. (14/15)
 into eq. (9) one obtains a third order equation for $\eta_1$ 
only\footnote{In {\sc M\"uller, Schmidt} (1983) the same 
vacuum  field equations were discussed
 for axially symmetric Bianchi type I models.
 They possess a four-parameter group of isometries, 
too,  and the essential field equation is also a 
third order equation for one function. 
The difference is that in the present case spherical
 symmetry implies an explicit coordinate dependence.}
 whose linearization reads 
\be%16
0 = \eta_1''' + 5\eta_1'' + \eta_1'(2 - e^{2x}) - \eta_1(8+ 4e^{2x}) \,  .
\ee
The solution of eq. (16) which is bounded for $x \to  - \infty$  
reads\footnote{A second
 solution is    $ \eta_1 = - 12 mr^{-4} $
leading to the Schwarzschild solution 
(this solution solves both the full and the linearized equations
 in accordance with the fact that it makes zero both sides 
of eq. (6)), and the third one can be obtained from them up
 to quadrature by usual methods.}
\be%17
\eta_1 = \gamma \cdot \left[ (3e^{-4x} + e^{-2x} )
 \sinh \, e^x  - 3 e^{-3x} \cosh \, e^x \right] \, .
\ee
A comparison with (12) gives $\gamma  = 15\epsilon l^{-3}$.

Now we insert this $\eta_1$  into eq.
 (15/11) and neglect
 again powers of $\eta_1$; then the metric reads
\bea%18
\beta =   5 \epsilon\left[l r^{-1} \sinh (r l^{-1})
 - \cosh (rl^{-1}) \right] \, , 
             \nonumber            \\
\lambda = \frac{5 \epsilon}{2 l} \cdot
\int_0^r \left[
 (l^2 z^{-2} - 1) \sinh (zl^{-1}) - l z^{-1} \cosh (z l^{-1})
   \right]
dz \, .
\eea
This linearization has (in the contrary to (1) and (12))
 the following property: to each $r_0> 0$ and $\Delta > 0$ there
 exists an $\delta > 0$ such that for $- \delta < \epsilon < \delta$ 
the relative error of the linearized solution (18) 
does not exceed $\Delta$ uniformly on the interval $0 \le  r \le  r_0$.

\section{Non-standard analysis}%4
In this section we will prove that in a neighbourhood of Minkowski 
space there do not exist any solutions. 
``Neighbourhood of Minkowski space''
 is in general a concept requiring additional 
explanations because of the large number of 
 different topologies discussed in literature. 
To make the above statement sufficiently strong 
we apply a quite weak topology here: given a $\delta > 0$
 then all space-times
 being diffeomorphic to Minkowski space and fulfilling
$ \vert R_{ij} R^{ij} \vert < \delta k^2$ 
form a 
neighbourhood about Minkowski space.

In (13) we have at $r =0$
 (independently of higher order terms in $r$) 
\be%19
R_{ij} R^{ij}
     = 100 k^2 \epsilon^2/3 \qquad {\rm  and} \qquad  R_{00} =  5k\epsilon
\, .
\ee
Therefore it holds: the one-parameter family of solutions
 being regular at $r = 0$ is invariantly characterized
 by the real 
parameter $\epsilon$, and a necessary condition for it to be within 
a neighbourhood of Minkowski space is that $\epsilon$
 lies in a neighbourhood of $0$.

Now we suppose $\epsilon$
 to be an infinitesimal number,  i.e., a positive number which is smaller 
than any positive real number. 
The mathematical theory dealing with such infinitesimals 
is called non-standard analysis, cf. {\sc Robinson} (1966). 
The clue is that one can handle non-standard 
numbers like real numbers. Further we need the so-called
{\it Permanence principle}  (= Robinson lemma): let $A(\epsilon)$
be an internal statement holding for all 
infinitesimals $\epsilon$. Then there exists a positive standard real 
$\delta > 0$ such that $A(\epsilon)$  holds for all $\epsilon$
 with   $0 < \epsilon < \delta$.
 The presumption ``internal"
  says, roughly speaking, that in the formulation 
of $A (.)$  the words ``standard''
 and ``infinitesimal" do not appear. For a more detailed explanation cf.
  the literature.

This permanence principle shall be applied as follows:  $A(\epsilon)$ 
is the statement: ``Take (12) as initial condition for eqs. (8-10)
 and calculate the corresponding metric (7/11). Then there exists 
an $r_0 \le l$  such that $R_{ij} R^{ij} \ge  k^2$ at
$r =   r_0$.''

\noindent
{\it Remark.} At this point it is not essential whether $r_0$  is 
a standard or an (infinitely large) non-standard number.

\noindent 
{\it Proof} of     $A(\epsilon)$  for infinitesimals
   $\epsilon$ 
  with   $\epsilon \ne 0$: the difference 
of (18) in relation to the exact solution is of the order $\epsilon^2$, 
i.e. the relative error is infinitesimally small. 
For increasing $r$, $\,   \vert \beta \vert $
 becomes arbitrary large, i.e. $(1 + \beta)^{-1}$ 
 becomes small and $R_{ij}R^{ij}$ 
 increases to arbitrarily large values, cf. eq. (18).
  Now take $r_0$ such that with metric (7/18)  $R_{ij}R^{ij}
\ge  2k^2 $
 holds. Then, for the exact solution, 
$R_{ij}R^{ij} \ge  k^2 $ 
   holds at $r = r_0$, because their 
difference was shown to be infinitesimally small.

Now the permanence principle tells us  that 
there exists a positive standard real $\delta> 0$
 such that for all $\epsilon$ with $0 < \vert \epsilon \vert  < \delta$   
the corresponding exact solution has a  
point $ r_0$  such that at $r=  r_0 \ $  
 $\   R_{ij}R^{ij} \ge  k^2 $ 
 holds. But ``$R_{ij}R^{ij} <  k^2 $''  is another necessary 
condition for a solution to lie within a neighbourhood of  Minkowski space.

\noindent 
{\it Remark.}
 Supposed this $r_0$ is an infinitely large non-standard number, then by 
continuity arguments 
also a standard (finite) number with the same property exists.

Now the statement is proved, 
but we have learned nothing about the actual value of the number $\delta$. 
Here,
 numerical calculations may help. 
They were performed as follows: the power series for the
 functions $\eta_i$
were calculated up to the $r^6$-term, then these functions taken at $x = -4 $
 (i.e. $r = 0.018 \ l$)  were used as initial conditions for  a Runge--Kutta 
integration of eqs. (8-10). 
We got the following result: firstly, for $\epsilon  = \pm 10^{-5}$
 and $r \le 10 \ l$, the 
relative difference between the
 linearized solution (17) and the numerical
 one is less than 2 per thousand. 
Secondly, for $0 < \vert \epsilon \vert \le 1$  the behaviour 
``$\beta \to 
   - \infty \cdot {\rm sgn} \epsilon$" is
 confirmed. That means, the statement made above keeps valid at
least for $\delta = 1$.

\noindent 
{\it Remark.}  For large values 
$\epsilon$ the power series for the $\eta_i$
 converge very slowly, 
and therefore  other methods
 would be necessary to decide about asymptotical flatness.

\section{Discussion  -- Einstein's particle programme}%5
Fourth order gravitational 
field equations could be taken as a field theoretical model 
of ordinary matter, the energy--momentum tensor of which is defined by
\be%(20)
\kappa T^\ast_{ij}= R_{ij} - \frac{1}{2} g_{ij} R   \, .
\ee
For our case one obtains at $r=  0 \ $ $\ T^\ast_{ij}$
 to represent an ideal fluid with the equation of state $p^\ast =\mu^\ast/3 $
  and (cf. eq. (19))
\be%21 
\mu^\ast (0) =     5 k \epsilon  \kappa^{-1} =
5 \epsilon /8\pi Gl^2 \, . 
\ee
Inserting $\vert \epsilon \vert \ge 1$
 and $l \le 1.3\cdot 10^{-13}$~cm into eq. (21), we obtain
\be%   (22)
\vert \mu^\ast(0) \vert \ge 1.5 \cdot 10^{53} \ {\rm g \ cm}^{-3} \, . 
\ee
Therefore it holds: if there
 exists a non-trivial static spherically symmetric
 asymptotically flat
 singularity-free solution of eq. (5) at all, then the
 corresponding particle would be a very massive one: 
its phenomenological energy density exceeds that of a neutron 
star by at least 40 orders of magnitude.

The resulting statement can be understood as follows: 
For a small curvature the 4th order corrections to Einstein's equations are 
small, too, and  the situation should not be very different from that one we 
know from Einstein's theory.

Now we want to refer 
to a problem concerning Schwarzschild coordinates: the transition 
from a general static spherically symmetric 
 line element to Schwarzschild coordinates is possible 
only in the 
case that the function ``invariant surface of the sphere $r =$ const. 
in dependence on its invariant radius" has 
not any stationary point. 
Here two standpoints are possible: either 
one takes this as a natural condition for a reasonable particle
 model or one allows coordinate singularities in (7) 
like ``$\beta \ge -1$, and $\beta = -1$  at single points" \footnote{For
 $\beta < -1$ one would obtain a cosmological model of 
Kantowski-Sachs type. Eqs. (2-11) remain unchanged for this case.}
 which require a special care. 
(The discussion made above is not influenced by this.)

The statement on the existence of solutions can be strengthened 
as follows: {\sc Fiedler, Schimming} (1983)
 proved that the solutions are  analytical
 in a neighbourhood of $r = 0$.
 Further, the differential equation is an analytical one and, 
therefore, in the 
subspace of singularity-free solutions they remain so in the limit 
$r \to \infty$.

Then, there exists only a finite or countably infinite set 
of values $\epsilon_n$  such that the corresponding solution becomes 
asymptotically flat (the question, whether this
 set is empty or not, shall be subject of further investigation.); 
furthermore, the  $\epsilon_n$  have no
 finite accumulation point. 
That means, there exists at most a discrete spectrum of solutions.

With  respect to this
 fact, we remark the 
following: as one knows, Einstein's theory is a covariant one. 
But besides this symmetry, it is
 homothetically invariant, too. That means, if $ds^2$  is
 changed to   $e^{2 \chi}ds^2$
 with constant $\chi$, then the tensor 
$R_{ij} - \frac{1}{2} g_{ij} R$
 remains unchanged,   whereas the
 scalars $R$ and $C$
 will be divided by  $e^{2 \chi}$
 and $e^{4 \chi}$
 resp. From this it follows: with one solution 
of Einstein's  vacuum equation one obtains by  homothetical  invariance 
just a one-parameter class 
of solutions. On the other hand, the sum $R + l^2 C$ 
has not such 
a symmetry and, therefore, 
one should not expect that a one-parameter family of 
solutions globally exists, and  
this is just in the 
scope of the particle programme where a definite particle's
  mass is wanted.

{\it Acknowledgement.}  Discussions with
 {\sc  R. John, U. Kasper} and R. {\sc Schimming} 
on 4th order field equations and with {\sc  J. Reichert}$^{\dag}$ 
 and {\sc H. Tuschik} on 
non-standard analysis are gratefully acknowledged. 
Further I want to thank W. {\sc Mai}
 and M. {\sc Schulz} for supporting the numerical calculations.

\section*{References}

\noindent
 BACH, R.: 1921, Math. Zeitschr. {\bf  9}, 110.

   \noindent
BORZESZKOWSKI, H. v.: 1981, Ann. Phys. (Leipz.) {\bf  38}, 239.

   \noindent
 BORZESZKOWSKI, H. v., TREDER, H.,
 YOURGRAU, W.: 1978, Ann. Phys. (Leipz.) {\bf  35}, 471.

   \noindent
 EINSTEIN, A.: 1921, Sitzungsber. AdW, Berlin, {\bf 1}, 261.

   \noindent
EINSTEIN, A., PAULI, W.: 1943, Ann. Math. {\bf 44}, 131.

   \noindent
    FIEDLER, B., SCHIMMING, R. :1983, Astron. Nachr. {\bf 304}, 221.

   \noindent
    M\"ULLER, V., SCHMIDT,  H.- J.; 1983, submitted to Gen. Rel. Grav.
\footnote{The correct source is: 
H.-J. Schmidt, V. M\"uller: On Bianchi type I vacuum
solutions in $R + R^2$ theories of gravitation II. The axially
symmetric anisotropic  case, Gen. Rel. Grav. {\bf  17} (1985) 
971-980. {\it (This footnote is not  in the original.)}}

   \noindent
PECHLANER, E., SEXL, R.: 1966, Comm. Math. Phys. {\bf  2}, 165.

   \noindent
    ROBINSON, A.: 1966, Non-standard analysis, North Holland, Amsterdam.

   \noindent
    STELLE, K. S.: 1978, Gen. Rel. Grav. {\bf 9}, 353.

   \noindent
    TREDER, H.: 1975, Ann. Phys. (Leipz.) {\bf 32}, 383.  

   \noindent
 TREDER, H.: 1977, p. 279 in: 75 Jahre Quantentheorie, ed. BRAUER, W., 
Akad.-Verlag  Berlin.

   \noindent
WEYL, H.: 1919, Ann. Phys. (Leipz.) {\bf  59}, 101.

   \noindent
    W\"UNSCH, V. : 1976, Math. Nachr. {\bf 73}, 37.

\medskip

\noindent
(Received 1984 March 20)

\medskip

\noindent 
{\small {In this reprint (done 
with the kind permission of the copyright owner) 
we removed only obvious misprints of the original, which
was published in Astronomi\-sche Nachrichten  
 under the  title ``On static spherical 
 symmetric solutions of the Bach-Einstein gravitational field equations'', 
 Astron. Nachr. {\bf 306} (1985) Nr. 2, pages 67 - 70;  
  Author's address that time:  
Zentralinstitut f\"ur  Astrophysik der AdW der DDR, 
1502 Potsdam--Babelsberg, R.-Luxemburg-Str. 17a.}}

\end{document}